\documentclass[%
twocolumn,
reprint,
showpacs,
 amsmath,amssymb,
 aps,
]{revtex4}

\usepackage{graphicx}
\usepackage{dcolumn}
\usepackage{bm}
\usepackage{lipsum}
\usepackage{color}
\usepackage{ulem}
\usepackage{ulem}

\begin{document}
\newcommand{\reviewstrike}[1]{\sout{#1}}

\preprint{XXXX}

\title{Dynamical properties of the honeycomb-lattice Iridates ${\rm Na_2IrO_3}$}

\author{
Takafumi Suzuki$^1$, Takuto Yamada$^1$, Youhei Yamaji$^2$, and Sei-ichiro Suga$^1$
}

\affiliation{${}^1$Graduate School of Engineering, University of Hyogo, Himeji 671-2280, Japan\\
${}^2$Quantum-Phase Electronics Center and Department of Applied Physics, University of Tokyo, Bunkyo-ku 113-0033, Japan}

\date{\today}

\begin{abstract}
We investigate the dynamical properties of ${\rm Na_2IrO_3}$.
For five effective models proposed for ${\rm Na_2IrO_3}$, 
we numerically calculate dynamical structure factors (DSFs) with an exact diagonalization method. 
An effective model obtained from \textit{ab initio} calculations explains inelastic neutron scattering experiments adequately.
We further calculate excitation modes based on linearized spin-wave theory. 
The spin-wave excitation of the effective models obtained by \textit{ab initio} calculations disagrees with the low-lying excitation of DSFs. 
We attribute this discrepancy to the location of ${\rm Na_2IrO_3}$ in a parameter space close to the phase boundary with the Kitaev spin-liquid phase.
\end{abstract}

\maketitle
%
%
%
\section{Introduction}
\label{sec:Introduction}
Magnetic properties in 4$d$ and 5$d$ transition metal compounds have attracted much attention in condensed matter physics. 
In some materials, such as ${\rm Na_2IrO_3}$, the energy scales of spin--orbit interactions, on-site coulomb interactions, and crystal fields compete with each other. The competition can produce unusual phases, including topological insulators \cite{Shitade2009} and Kitaev spin liquids \cite{JChaloupka2010}.

In ${\rm Na_2IrO_3}$, ${\rm Ir^{4+}}$ ions can be expressed as an isospin with a total angular momentum of $1/2$ \cite{GJackeli2009}. 
We refer to this isospin as "spin" hereafter. 
${\rm IrO_6}$ octahedrons in ${\rm Na_2IrO_3}$ form a planar structure parallel to the $ab$ plane, and ${\rm Ir^{4+}}$ ions constitute a honeycomb lattice~\cite{GJackeli2009,JChaloupka2010}.  
In addition, ${\rm IrO_6}$ octahedrons are connected by sharing the oxygen atoms on the edges, making the Ir-O-Ir bond angle nearly $90^{\circ}$. 
This causes three kinds of anisotropic interactions between ${\rm Ir^{4+}}$ ions depending on the bonding-path direction. 
${\rm Ir^{4+}}$ ions can also interact via direct overlap of their orbitals. 
Thus, both Kitaev and Heisenberg interactions occur between ${\rm Ir^{4+}}$ ions \cite{JChaloupka2010}, leading to the Kitaev--Heisenberg model.

${\rm Na_2IrO_3}$ undergoes a magnetic phase transition to a zigzag antiferromagnetic order at $T_{\rm N} \sim 15 $ K~\cite{SKChoi2012,FYe2012}. 
In a typical Kitaev--Heisenberg model, where ferromagnetic Kitaev and antiferromagnetic Heisenberg interactions are summed for nearest neighbor pairs, 
the zigzag order is not stabilized; thus,  
several models have been proposed to explain the zigzag ordering~\cite{IKimchi2011,SKChoi2012,JChaloupka2013,VMKatukuri2014,YYamaji2014,YPSizyuk2014}.  
Some models~\cite{IKimchi2011,JChaloupka2013,YYamaji2014} have succeeded in explaining the temperature dependence of thermodynamic quantities, such as the specific heat and magnetic susceptibility. 
In discussing interaction parameters, \textit{ab initio} calculations are particularly powerful~\cite{RComin2012,YYamaji2014,VMKatukuri2014,YPSizyuk2014}.
However, there are considerable differences between the estimated parameters because they are sensitive to the approximations used in the calculations. 
Therefore, there is still controversy surrounding suitable models for ${\rm Na_2IrO_3}$.

The dynamical properties of ${\rm Na_2IrO_3}$ have been investigated by inelastic neutron scattering (INS) experiments~\cite{SKChoi2012}, and 
a linearized spin-wave analysis has explained the low-lying excitations observed in the experiments~\cite{SKChoi2012,JChaloupka2013}.
However, the proposed parameters differed between the studies.
Choi $et$ $al$.~\cite{SKChoi2012} discussed the importance of the long-range interactions, 
whereas Chaloupka $et$ $al$.~\cite{JChaloupka2013} proposed the other scenario in which the signs of the Kitaev and Heisenberg terms play a key role.
In addition, the \textit{ab initio} calculations indicated that ${\rm Na_2IrO_3}$ is located close to the phase boundary with the Kitaev spin-liquid phase \cite{YYamaji2014,VMKatukuri2014}. 
If the system is located close to the phase boundary,
degenerate low-lying excitations from the magnetic frustration caused by the dominant Kitaev couplings may make
conventional spin-wave theory invalid.
Therefore, it is important to investigate the dynamical properties of ${\rm Na_2IrO_3}$ by using a method that does not depend on an approximation.

In this paper, we numerically investigate the dynamical properties of ${\rm Na_2IrO_3}$ by an exact diagonalization method. 
We focus on dynamical structure factors (DSFs) of five effective models proposed for ${\rm Na_2IrO_3}$~\cite{SKChoi2012,JChaloupka2013,YYamaji2014,VMKatukuri2014,YPSizyuk2014}.
The DSFs provide magnetic excitations, which can be measured in INS experiments.  
We compare our numerical results with the experimental results for the powder samples of this compound~\cite{SKChoi2012} and discuss the suitability of the models. 
To examine the low-lying excitations, we study excitation modes further by a linearized spin-wave analysis. 
The spin-wave excitations of the models obtained by the \textit{ab initio} calculations disagree with the low-lying excitations of the DSFs. 
We attribute this discrepancy to the location of ${\rm Na_2IrO_3}$ close to the phase boundary with the Kitaev spin-liquid phase.
Indeed, for an {\it ab initio} model~\cite{YYamaji2014}, we confirm a double peak structure in the specific heat, 
which can be a probe to observe the fractionalization of quantum spins in the Kitaev spin-liquid phase~\cite{JNasu2015}.

\begin{figure}[htb]
  \begin{center}
     \includegraphics[width=\linewidth]{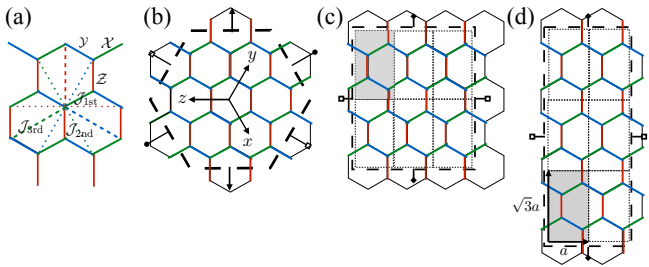}
\caption{\label{fig1} (Color online) (a)
Honeycomb lattice model. Red, blue, and green lines denote ${\mathcal Z}$-, ${\mathcal X}$-, and ${\mathcal Y}$- bonds, respectively. 
Dotted (dashed) lines represent the second (third) neighbor interactions. The colors of these lines are the same as those for the corresponding nearest neighbor lines. 
Different lattice geometries for $N=24$ in (b) the $\pi/3$-rotational symmetric case (${\rm C_3}$), (c) 3 $\times$ 2 cells (3$\times$2), and (d) 2 $\times$ 3 cells (2$\times$3). 
The $x$, $y$, and $z$ axes in (b) correspond to the orthogonal axes for spin operators that are defined from the 5$d$ $t_{2g}$-orbitals of ${\rm Ir^{4+}}$~\cite{YYamaji2014} and the honeycomb plane is perpendicular to the $(1,1,1)$ direction.
Black bold dashed lines enclose the 24 sites, and shaded rectangles indicate four sublattice unit cells. 
In (d), a schematic of the zigzag order is presented. For all geometries, we apply periodic boundary conditions and boundaries with the same symbol are connected.}
 \end{center}
\end{figure}

The layout of this paper is as follows.
In Sec. \ref{sec:Model and Method}, we introduce five effective models proposed for ${\rm Na_2IrO_3}$.
We compute the DSFs by the numerical exact-diagonalization method to discuss the low-lying excitations of the five models.
In Sec. \ref{ResultsDiscussion}, we discuss the low-lying excitations of the DSFs comparing with INS results for powder samples~\cite{SKChoi2012}.
To capture the properties of the low-lying excitations, we calculate spin-wave excitations.
We find that the spin-wave excitations fail to explain the low-lying excitations of the DSFs when the model is located nearby the Kitaev spin liquid phase.
Finally, we conclude the discussion in Sec. \ref{Discussions}.

\section{Model and Method}
\label{sec:Model and Method}
%
%
\subsection{Effective models for ${\rm Na_2IrO_3}$}
\begin{table*}[htb]
\begin{center}
  \caption{\label{table1} Interactions and effects of trigonal distortions in the five proposed models. 
`$\surd$' (`$-$') indicates that the corresponding term is included (not included).
`$\ast$' indicates that it was present, but the value was not estimated.
F (AF) denotes ferromagnetic (antiferromagnetic) Kitaev interactions. NN (NNN) denotes next-(next-) nearest-neighbor interactions.
LSW: linearized spin-wave theory. DFT: density-functional theory. QC: wave-function based quantum chemistry.}
  \begin{tabular}{lcccc}
  \hline \hline 
  Model & Long-range interaction & Trigonal distortion & Kitaev interaction & Method\\ \hline 
  Model I:  \cite{SKChoi2012} & $\surd$ &  $-$ &  $-$ & LSW\\
  Model II: \cite{JChaloupka2013} & $-$ & $-$ & AF  & LSW\\ 
  Model III: \cite{YYamaji2014} &  $\surd$ & $\surd$ & F & \textit{ab initio} [DFT]\\
  Model IV: \cite{VMKatukuri2014} &  $\ast$ & $\surd$ & F & \textit{ab initio} [QC]\\
  Model V:  \cite{YPSizyuk2014} & $\surd$ & $-$ & F for NN and AF for NNN  & \textit{ab initio} [DFT]
    \\ \hline \hline
  \end{tabular}
\end{center}
\end{table*}
We consider a generalized form of Kitaev--Heisenberg models on a honeycomb lattice. 
The Hamiltonian is given as
\begin{eqnarray}
\label{Ham1}
{\mathcal H}=&\displaystyle\sum_p \sum_{\Gamma_p} \sum_{\langle ij \rangle \in \Gamma_p} J_{\Gamma_p} {\boldsymbol S}_i \cdot {\boldsymbol S}_j \nonumber\\
+&\displaystyle\sum_p \sum_{\Gamma_p} \sum_{\langle ij \rangle \in \Gamma_p} \left[ K_{\Gamma_p}{S_i}^{\gamma}{S_j}^{\gamma} + D_{\Gamma_p} ({S_i}^{\alpha}{S_j}^{\beta}  + h.c.) \right],
\end{eqnarray}
where $\Gamma_p$ is the bond indices depending on the direction (${\mathcal X}$, ${\mathcal Y}$, ${\mathcal Z}$) of the $p$-th neighbor pairs 
and $\alpha$, $\beta$, and $\gamma$ are indices for $S=1/2$ SU(2)-spin components and take $x$, $y$, or $z$ cyclicly. 
Each assignment of the indices depends on the bond direction; for example, $\gamma=z$, $\alpha=x$ and $\beta=y$ for the ${\mathcal Z}$ bond direction and so on. 
From the symmetry of the crystal structure for Na$_2$IrO$_3$, the highly generalized form of the Hamiltonian reads
\begin{eqnarray}
\label{Ham2}
{\mathcal H}=
\displaystyle\sum_p \sum_{\Gamma_p} \sum_{\langle ij \rangle \in \Gamma_p} \sum_{\mu,\nu=x,y,z} S^{\mu}_i \hat{\mathcal J}^{\mu \nu}_{\Gamma_p} S^{\nu}_j,
\end{eqnarray}
where the exchange coupling between $i$ and $j$ sites on the bond ${\Gamma_p}$ is given by a $3\times3$ matrix $\hat{\mathcal J}^{\mu\nu}_{\Gamma_p}$. 
In this paper, we focus on the five models~\cite{SKChoi2012,JChaloupka2013,YYamaji2014,VMKatukuri2014,YPSizyuk2014} shown in Table \ref{table1}. 
We can summarize three key factors in the five models: long-range interactions, trigonal distortions, and signs of the Kitaev interaction.
In Models I and II, the interactions were evaluated by a spin-wave analysis so as to reproduce the low-lying excitations observed in the INS experiments and the temperature dependence of thermodynamic quantities~\cite{SKChoi2012,JChaloupka2013}.  
In contrast, the interactions in Models III--V were estimated from the \textit{ab initio} calculations~\cite{YYamaji2014,VMKatukuri2014,YPSizyuk2014}. 
In Models III and V, interactions were estimated from the density-functional-theory calculations. 
Note that there are technical differences between Models III and V in evaluating the tight-binding models and the consequent effective spin model. 
In contrast,  in Ref. ~\cite{VMKatukuri2014}, Katukuri and co-workers employed \textit{ab initio} techniques from wave-function-based quantum chemistry and 
proposed several parameter sets for the coupling constants.
We adopt the nearest-neighbor interactions that were used in the right phase diagram in Fig. 2 of Ref. \cite{VMKatukuri2014} as Model IV.
For the second and third neighbor Heisenberg interactions, we adopt the middle values between the proposed range for explaining the experimental Curie-Weiss temperature~\cite{VMKatukuri2014}.  
Details of the interaction parameters for the five models are summarized in Tables \ref{table2} and \ref{table3}.

\begin{table}
 \caption{\label{table2} Coupling constants except for Model III. The positive (negative) number corresponds to the antiferromagnetic (ferromagnetic) interactions. 
 For Models I, II, IV, and V, the Hamiltonian can be expressed by the expression (\ref{Ham1}) with $D^{\rm 2nd}=D^{\rm 3rd}=0$.  The energy unit is meV.}
\begin{tabular}{lcccccccc}
\hline \hline 
            & $J^{\rm 1st}$ & $K^{\rm 1st}$  &  $D^{\rm 1st}$  & $J^{\rm 2nd}$& $K^{\rm 2nd}$ & $J^{\rm 3rd}$ & $K^{\rm 3rd}$ \\
            \hline
Model I:~\cite{SKChoi2012} & 4.17 & 0  & 0 & 3.25 & 0 & 3.75 & 0\\
Model II:~\cite{JChaloupka2013} & -4.0 & 21.0  & 0 & 0 & 0 & 0 & 0\\ 
Model IV:~\cite{NOTE2}  & 3 & -17.5 & -1 & 4.5 & 0 & 4.5 & 0\\
Model V:~\cite{YPSizyuk2014} & 5.8 & -14.8  & 0 & -4.4 & 7.9 & 0 & 0\\ 
\hline
\\
\end{tabular}
\end{table}

\begin{table*}
 \caption{\label{table3} Coupling constants for Model III. Model III is expressed by the Hamiltonian (\ref{Ham2}). 
 The energy unit is meV.  `-' means that the coupling is not included. }
\begin{tabular}{cccccccccccccccccccccccccccccccccccc}
\hline \hline 
  & \multicolumn{4}{c}{$J^{\rm 1st}_{\mathcal X}$} & \multicolumn{4}{c}{$J^{\rm 1st}_{\mathcal Y}$} & \multicolumn{3}{c}{$J^{\rm 1st}_{\mathcal Z}$}   &  \multicolumn{4}{c}{$J^{\rm 2nd}_{\mathcal X}$} & \multicolumn{4}{c}{$J^{\rm 2nd}_{\mathcal Y}$} & \multicolumn{3}{c}{$J^{\rm 2nd}_{\mathcal Z}$}  &  \multicolumn{4}{c}{$J^{\rm 3rd}_{\mathcal X}$} & \multicolumn{4}{c}{$J^{\rm 3rd}_{\mathcal Y}$} & \multicolumn{3}{c}{$J^{\rm 3rd}_{\mathcal Z}$} \\ 
  & $x$ & $y$ & $z$ & & $x$ & $y$ & $z$ &  & $x$ & $y$ & $z$ & $x$ & $y$ & $z$ & & $x$ & $y$ & $z$ &  & $x$ & $y$ & $z$ & $x$ & $y$ & $z$ & & $x$ & $y$ & $z$ &  & $x$ & $y$ & $z$   \\ 
   \hline 
                                                    $x$& -23.9 & -3.1  & -8.4  & & 2.0 & -3.1   & 1.8  &  & 4.4 & -0.4 & 1.1  & \multicolumn{4}{c}{} & \multicolumn{4}{c}{} & -0.8 & 1.0 & -1.4 & 1.7 & 0 & 0 & & 1.7 & 0 & 0 & & 1.7 & 0 & 0\\
                                                    $y$& -3.1   & 3.2   & 1.8    & & -3.1& -23.9 & -8.4 &  & -0.4 & 4.4 & 1.1  & \multicolumn{4}{c}{-} & \multicolumn{4}{c}{-} & 1.0 & -0.8 & -1.4 & 0 & 1.7 & 0 & & 0 & 1.7 & 0 & & 0 & 1.7 & 0\\
                                                     $z$& -8.4  & 1.8   & 2.0    & & 1.8 & -8.4   & 3.2  &  & 1.1 & 1.1 & -30.7& \multicolumn{4}{c}{} & \multicolumn{4}{c}{} & -1.4 & -1.4 & -1.2 & 0 & 0 & 1.7 & & 0 & 0 & 1.7 & & 0 & 0 & 1.7  \\  \hline
\hline
\end{tabular}
\end{table*}

\subsection{Dynamical structure factors}

For the five models presented in the previous subsection, we calculate the DSF, $S({\boldsymbol Q}_i,\omega)$, for the system size $N=24$ with three different lattice geometries shown in Figs. \ref{fig1}(b)--\ref{fig1}(d).
The DSF at a zero temperature is defined as
\begin{eqnarray}
S^{\mu \nu}({\boldsymbol Q},\omega) \equiv -\frac{1}{\pi} \lim_{\epsilon \rightarrow +0} {\rm Im} \langle \phi_0 | \frac{\hat{S}^{\mu \dagger}_{\boldsymbol Q}  \hat{S}^{\nu}_{\boldsymbol Q}}{\omega+E_0+i\epsilon-{\mathcal H}} |\phi_0\rangle,
\end{eqnarray}
where $\phi_0$ is the ground state of ${\mathcal H}$ with the energy $E_0$ and $\hat{S}^{\nu}_{\boldsymbol Q} = \frac{1}{N}\sum_{\boldsymbol r} {S}^{\nu} \exp(-i{\boldsymbol Q}\cdot{\boldsymbol r})$. 
$\phi_0$ and $E_0$ are calculated by the Lanczos method, and then $S^{\mu \nu}({\boldsymbol Q},\omega)$ is obtained by a continued fraction expansion~\cite{EGagliano1987,EDagotto1994}.

In general, when the off-diagonal elements of $\hat{\mathcal J}^{\mu \nu}_{\Gamma_p}$ are non-zero, the DFS $S^{\mu \nu}({\boldsymbol Q},\omega)$ is allowed to have non-zero off-diagonal elements.
The contribution from the off-diagonal elements in the DFSs is expected to be proportional to the off-diagonal spin correlation.
The amplitude of the spin correlation is also proportional to the absolute values of the matrix element of $\hat{\mathcal J}^{\mu \nu}_{\Gamma_p}$.
In the five models presented in the previous section, the sum of all diagonal elements in $\hat{\mathcal J}^{\mu \nu}_{\Gamma_p}$ is larger than the remaining each elements. 
Therefore, we consider that the sum of diagonal elements, namely $S({\boldsymbol Q},\omega)=\sum_{\mu}S^{\mu \mu}({\boldsymbol Q},\omega)$, mainly contributes to the scattering intensity.

The magnetic excitations of ${\rm Na_2IrO_3}$ have been investigated by INS experiments for the powder samples~\cite{SKChoi2012}.
Therefore, the scattering intensity observed in the experiments is averaged with respect to the wave vectors and the scattering directions.
In order to compare the experimental results with numerical results, we consider the averaged intensity defined as
${\mathcal I}(|{\boldsymbol Q}|,\omega)=\sum_{|{\boldsymbol Q}|={\boldsymbol Q}_i} S({\boldsymbol Q}_i,\omega)$, 
where $S({\boldsymbol Q}_i,\omega)=\sum_{\mu=x,y,z} S^{\mu \mu}({\boldsymbol Q}_i,\omega)$ and the sum for ${{\boldsymbol Q}_i}$ runs all possible wave vectors inside the first Brillouin zone.
Thus the averaged intensity, ${\mathcal I}(|{\boldsymbol Q}|,\omega)$, is characterized as a function of the distance from $\Gamma$ point.

\section{Results}
\label{ResultsDiscussion}
\subsection{Ground states of five models}
\begin{figure*}[thb]
  \begin{center}
   \includegraphics[width=\linewidth]{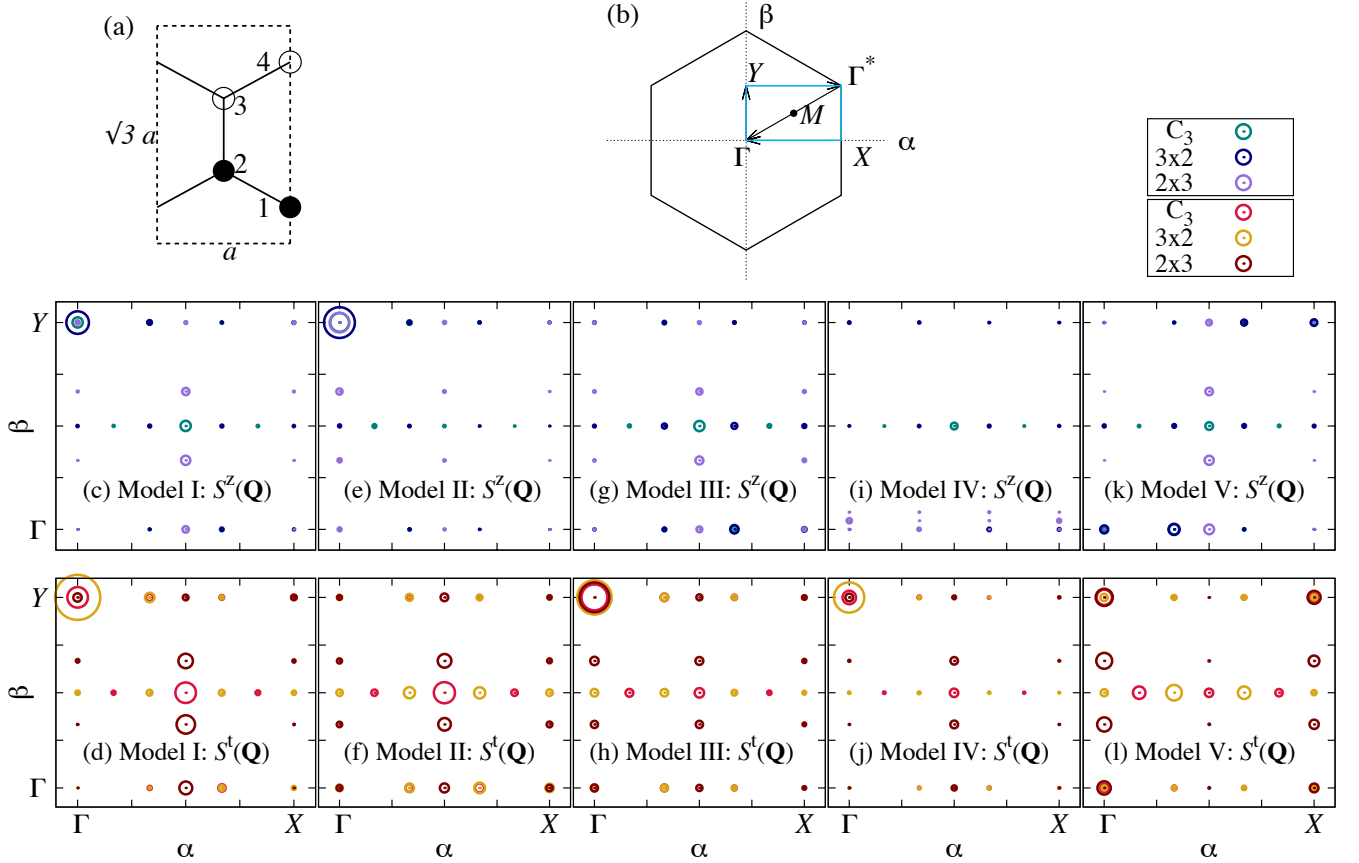}
\caption{\label{fig2} (Color online) (a) Four sublattice unit cells and schematic spin configuration of the Z-type collinear zigzag order. 
Solid (open) circles represent up (down) spins and the dotted rectangle corresponds to the shaded ones in Fig. \ref{fig1}.
(b) Reciprocal vectors $\alpha$ and $\beta$ for the four sublattice unit cells in (a). 
(c)--(l) SSFs for Models I--V for $N=24$ with three lattice geometries in Figs. \ref{fig1}(b)--\ref{fig1}(d). 
The area of each circle is proportional to the amplitude. 
Upper (lower) panels are results for the longitudinal (transverse) component. }
  \end{center}
\end{figure*}

Figure \ref{fig2} shows the results for the static structure factors (SSFs) for the five models.
The longitudinal and transverse elements of the SSFs are defined as $S^z({\boldsymbol Q})=\int^{\infty}_{0} S^{zz}({\boldsymbol Q},\omega) d\omega$ and $S^t({\boldsymbol Q)}=\int^{\infty}_{0} (S^{xx}({\boldsymbol Q},\omega)+S^{yy}({\boldsymbol Q},\omega)) d\omega$, respectively.
Although the amplitude of the SSFs depends on the lattice geometry, the largest peak appears at the Y and M points in Models I--V.
The results indicate that the stable ground state of the five models is a zigzag order.
This is consistent with the experimental results~\cite{SKChoi2012,FYe2012}.

Because we expect that the results of the $\pi/3$-rotational-symmetric lattice, which is labeled ${\rm C_3}$ in Fig. \ref{fig2}, 
well describe the properties at the thermodynamic limit, we focus on the results for the ${\rm C_3}$ case.
The ground states, except for in Model III, are characterized by a zigzag spin configuration, 
although the "type" discussed below cannot be assigned because of the C$_3$ three-fold rotational symmetry.
In Model III, the largest peak appears at the Y point in the transverse component. 
This means that the spin correlation on the ${\mathcal Z}$ bond is antiferromagnetic, 
and thus the ground state exhibits Z-type zigzag order~\cite{YYamaji2014}.  
Moreover each magnetic moment of the ground state points to a direction in the plane perpendicular to the $z$ axis [Fig. \ref{fig1}(b)].

\subsection{Powder averaged results}
\begin{figure*}[htb]
  \begin{center}
   \includegraphics[width=\linewidth]{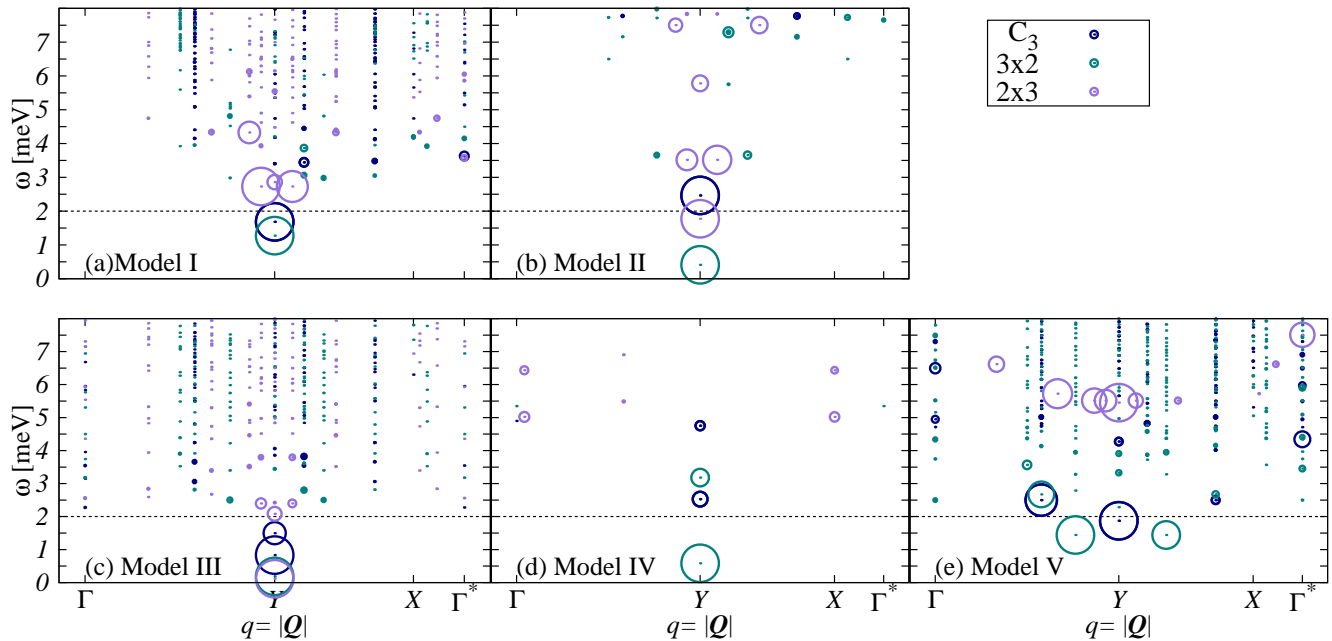}
\caption{\label{fig3} (color online) 
Averaged intensity, ${\mathcal I}(| {\boldsymbol Q} |, \omega)$, for $N=24$ with three lattice geometries. 
The area of each circle is proportional to the scattering intensity. 
All results are normalized by the largest value for each data set. 
Dotted lines are the low-energy cutoff at about $2$ meV in INS experiments \cite{SKChoi2012}. 
The horizontal axis corresponds to distance from the $\Gamma$ point. 
The labels on the horizontal axis denote the corresponding distance from the $\Gamma$ point to the label~\cite{comment1}.}
  \end{center}
\end{figure*}

Figure \ref{fig3} shows the numerical results for the averaged intensity, ${\mathcal I}(|{\boldsymbol Q}|,\omega)$.
We compare ${\mathcal I}(|{\boldsymbol Q}|,\omega)$ of the five models with the INS results for the powder samples shown in Fig. 3 of Ref. \cite{SKChoi2012}. 
The INS experiments \cite{SKChoi2012} show that there are three characteristic features for $T<T_{\rm N}$:
(i) a sharp lower boundary for the scattering intensity below 4 meV between the ${\Gamma}$ point and Y point (except for $|{\boldsymbol Q}|\sim 0$ and $\omega \lesssim 2$ meV, where the reported data have been lacked),
(ii) strong scattering peaks at close to 4 meV between the Y point and $\Gamma^{*}$ point, 
and (iii) no strong intensities at the $\Gamma^{*}$ point~\cite{comment1}.

First, we discuss the excitation boundary. 
The excitation boundary begins at ~4 meV nearly midway between the ${\Gamma}$ point and Y point,
 and the boundary energy decreases towards the Y point \cite{SKChoi2012}. 
The low-lying excitation energy in the related region for Models I--IV draws the convex curvature and it agrees with the experimental results.
In contrast, the low-lying excitation energy in Model V also increases from the Y point.
However, it exhibits a strong lattice-geometry dependence and the convex curvature can not be observed. 
The maximum energies of such curvature for Models I--IV locate at nearly midway between the ${\Gamma}$ point and Y point and 
are about 3.8 meV, 6.4 meV, 2.6meV, and 5.3 meV, respectively. 
Therefore, the agreement is excellent for Model I.
Since Model I was estimated to explain the excitation boundary by the linearized spin-wave theory, the agreement is quite natural.
Despite the good agreement, Model I is inadequate for describing ${\rm Na_2IrO_3}$, because it does not include the Kitaev interactions. 
It is unlikely that there is no Kitaev term, when we consider the interaction path between ${\rm Ir^{4+}}$ ions in ${\rm Na_2IrO_3}$. 

%

Secondly, we focus on the peaks below $T=T_{\rm N}$ described in the experimental feature (ii).
When we compare the experimental results for $T<T_{\rm N}$ with those for $T>T_{\rm N}$, 
we observe scattering intensities derived from the magnetic ordering around 4 meV between the Y point and $\Gamma^{*}$ point. 
Our numerical results reproduce these scattering intensities, except for Models II and IV. 
However, in Models I and V, relatively large peaks appear around $\sim 3.8$ meV at the $\Gamma^{*}$ point.  
The presence of such peak contradicts the experimental feature (iii).
Thus, Model III is also considered as a good candidate for the model that explains the experimental features (i), (ii), and (iii). 
Therefore, among the five models, Model III is the most suitable for explaining the INS experiments for ${\rm Na_2IrO_3}$. 
However, the low-lying excitation of the DSFs  in Model III appears slightly lower than the experimental results~\cite{SKChoi2012}.
Thus, further examination of the second and third neighbor interactions is desirable to improve the accuracy of the theoretical prediction.

\begin{figure*}[htb]
  \begin{center}
   \includegraphics[width=\linewidth]{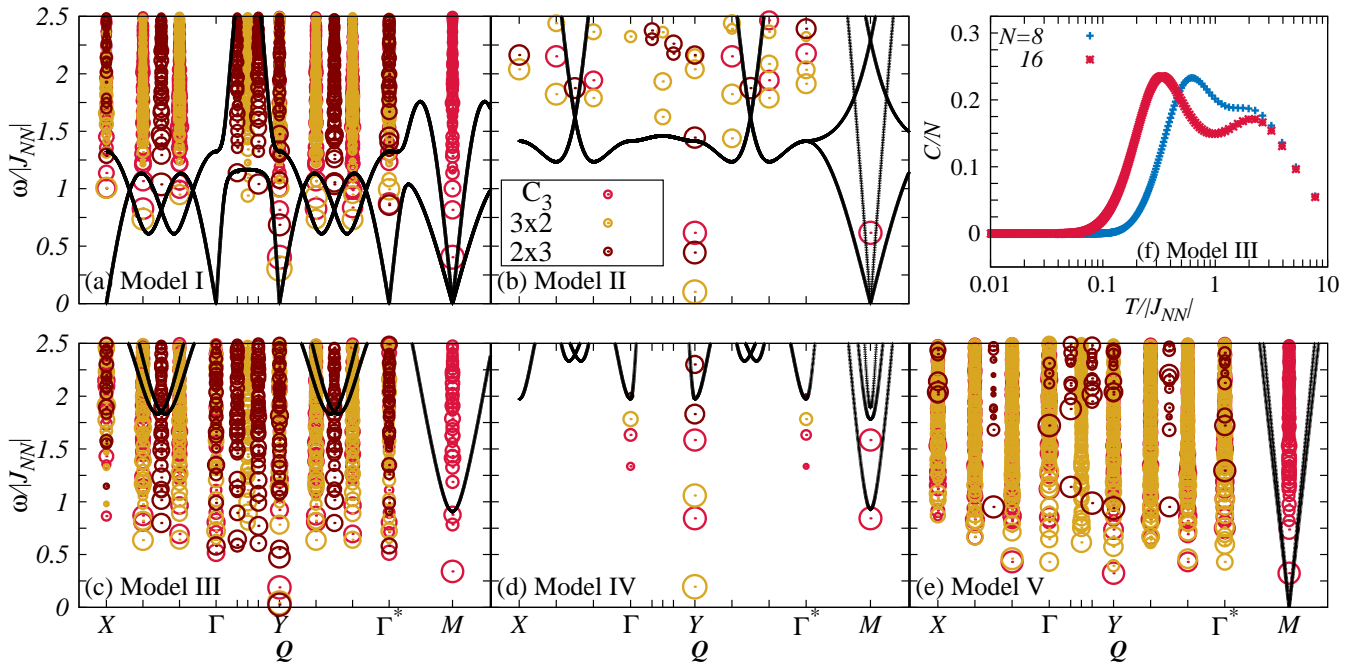}
\caption{\label{fig4} (Color online) 
DSFs, $S({\boldsymbol Q}_i,\omega)$, for $N=24$. Results for different lattice geometries are included.
The area of each circle corresponds to the logarithmic amplitude of the scattering intensity. 
The solid curves are dispersion curves estimated from the linearized spin-wave analysis. 
The excitation energy is normalized by the largest Heisenberg interaction $|J_{NN}|$ in each Model;
$|J_{NN}|$=4.17 meV, 4.0 meV, 4.4 meV, 3.0 meV, and 5.8 meV for Models I--V, respectively.
The horizontal axis in (a)--(e) runs along the arrows shown in Fig.~\ref{fig2} (b). 
The excitation energy is normalized.
(f) Temperature dependence of specific heat for Model III numerically obtained by a diagonalization method. Red (blue) symbols are the results for $N=16$ (8).}
  \end{center}
\end{figure*}
%

\subsection{Low-lying excitations}
Next, we discuss characteristics of the low-lying excitations of the DSFs, $S({\boldsymbol Q}_i,\omega)$.
We compare the DSFs with the spin-wave excitation modes for the four sublattice unit cells [Figs. \ref{fig4}(a)--\ref{fig4}(e)]. 
In the linearized spin-wave calculations for Models I, II, and V, we assume the Z-type collinear zigzag configuration shown in Fig. \ref{fig2}(a).
We examine the initial state of the spin-wave analysis for Model V, namely the ground state for the classical spin model.
Although Model V includes the long-range interactions expressed by the Kitaev--Heisenberg term,
the ground state analysis for classical spins indicates that the collinear zigzag order is still favored.
For Models III and IV, we obtain a collinear zigzag spin configuration for the ground state from the classical spins analysis. 
However the zigzag order parameter, which is defined as ${\boldsymbol M}={\boldsymbol S}_1+{\boldsymbol S}_2-{\boldsymbol S}_3-{\boldsymbol S}_4$ for four spins in the unit cell in Fig. \ref{fig2}(a), 
indicates that they are in a different direction from those for Models I, II, and V. 
This tilting of the order parameter results in the spin gap at the M point.

We turn our attention to the results for Model III.
The spin-wave excitations of Model III appear in the higher energy region above $\omega/J_{NN} \sim 1$ with an energy gap at the M point.
The low-lying excitation energies of the spin-wave modes at other ${\boldsymbol Q}$s are a few-times higher than those of the DSFs.
Thus, in Model III, the conventional spin-wave theory clearly fails to describe the low-lying excitations of ${\rm Na_2IrO_3}$.

A similar breakdown of the conventional linearized spin-wave picture for the low-lying excitations of the DSFs is also observed in Models IV and V, 
where the parameters are also estimated from \textit{ab initio} calculations. 
In Model V, except for the M point, the low-lying excitation of the spin-wave mode at each ${\boldsymbol Q}$ also appears in the high-energy region.
We here comment on the high-energy excitations in the DSFs. 
We observe some poles in the DSFs at similar energies to the spin-wave excitations, although those are not shown in Fig. \ref{fig4} owing to the large discrepancy in the energy scale.
The observed breakdown of the spin-wave excitations indicates that the low-lying excitations of the DSFs in Models III--V are apparently different from the free magnon excitations 
and can be attributed to precursors to deconfined Majorana excitations, which are composed of itinerant Majorana fermions and $Z_2$ gauge fields~\cite{JNasu2014,JNasu2015}.

\section{Discussions}
\label{Discussions}
The spin-wave picture fails to explain the low-lying excitations in \textit{ab initio} models for ${\rm Na_2IrO_3}$. 
This becomes significant when the model is close to the boundary of the spin-liquid phase, as for Model III~\cite{YYamaji2014}.
Recently, a similar breakdown of the spin-wave picture has been discussed in INS measurements for a potential Kitaev material, $\alpha$-${\rm RuCl_3}$ \cite{ABanerjee2015}. 
This material is near the Kitaev spin-liquid phase boundary~\cite{Plump2014,Sandi2015,Sears2015,Majumdar2015,Luke2015,Kubota2015}. 
Close to the Kitaev spin-liquid phase boundary, 
we expect that a precursor of the fractionalization of quantum spins into itinerant Majorana fermions and $Z_2$ gauge fields \cite{JNasu2014,JNasu2015} will be observed and may break the spin-wave picture for low-lying excitations.
The double peak structure in the specific heat can be used as a probe to observe the emergence of the fractionalization in the Kitaev spin liquid for the current two-dimensional models~\cite{JNasu2015}. 
In Model III, we confirm the double peak structure in the temperature dependence of the specific heat [Fig.~\ref{fig4}(f)].
Thus, the present results demonstrate that ${\rm Na_2IrO_3}$ is located close to the Kitaev spin-liquid phase boundary, if Model III is the most suitable.

The trigonal distortion is also important for the `distance' from the Kitaev spin-liquid phase in Model III~\cite{YYamaji2014}. 
We believe that systematic comparisons with the energy scale of the low-lying excitation of the DSFs, the peak separation of the specific heat, and the trigonal interaction may clarify discussions of the Majorana physics in experimental observations.

\section*{Acknowledgments}
\label{ackno}
We thank M. Imada, N. Kawashima, T. Okubo, and T. Tohyama for fruitful discussions.  
This work was supported by JSPS KAKENHI (Grants No. 25287104, No. 15K05232, and No. 15K17702). 
We acknowledge the computational resources of the K computer provided by the RIKEN Advanced Institute for Computational Science through the HPCI System Research Project (Projects ID: hp120283 and ID: hp130081).
We also acknowledge the numerical resources provided by the ISSP Supercomputer Center at University of Tokyo  
and the Research Center for Nano-micro Structure Science and Engineering at University of Hyogo.



\begin{thebibliography}{23}
\bibitem[{com()}]{comment1}
\bibinfo{note}{Y and $\Gamma^{*}$ denote the distance of the corresponding
  points from the $\Gamma$ point and they are not the real positions in the
  reciprocal space there.}
\end{thebibliography}
\end{document}